# ANALYSIS OF NET MIGRATION BETWEEN THE PORTUGUESE REGIONS

**Vítor João Pereira Domingues Martinho**

Unidade de I&D do Instituto Politécnico de Viseu
Av. Cor. José Maria Vale de Andrade
Campus Politécnico
3504 - 510 Viseu
**(PORTUGAL)**
**e-mail:** vdmartinho@esav.ipv.pt

**ABSTRACT**

This work aims mainly to present a project of research about the identification of the determinants that affect the mobility of labor. The empirical part of the work will be performed for the NUTS II and NUTS III of Portugal, from 1996 to 2002 and for 1991 and 2001, respectively (given the availability of statistical data). As main conclusion it can be said, for the NUTS II (1996-2002), which is confirmed the existence of some labor mobility in Portugal and that regional mobility is mainly influenced positively by the output growth and negatively by the unemployment rates and by the weight of the agricultural sector. NUTS III level (1991 and 2001) is something similar, but with this level of spatial disaggregation (and in this period) the basic equipment (amenities), particularly in terms of availability of housing, are the main determinants of migration.

## 1. INTRODUCTION

There are many authors who have dedicated themselves to issues of labor mobility, with very different theoretical assumptions, trying to investigate how these issues or do not explain the regional differences (1)(Martinho, 2006). For example, the authors associated with the Neoclassical theory, as (2)Solow (1956), consider that the tendency is, for the labor mobility, to alleviate, in the medium and long term, the regional disparities. This, because these authors consider the mobility of factors as a function of wages and the supply of resources as exogenous. Thus, what determines the mobility factor is their compensation.

On the other hand, the works in line with the Keynesian theory, such as (3)Myrdal (1957) and (4)Kaldor (1966), among others, argue that the trend is for labor mobility accentuate regional differences, these authors argue that because the existence of growth processes with circular and cumulative causes. This comes from assuming the existence of increasing returns to scale, to admit endogenous factors and to consider forces of demand (especially in foreign demand) as the main determinants of the growth process. Thus, factor mobility is a function of the forces of demand and employment moves to where demand is strong.

More recently, authors associated with the New Economic Geography, as (5)Fujita et al. (2000), among others, are also in favor of the labor mobility accentuates regional disparities. This derivative, as well as in the Keynesian theory (although with different assumptions), to assume the existence of growth processes with circular and cumulative causes. The assumptions for the New Economic Geography are microeconomic and have much to do with transportation costs "iceberg" and the existence of perfect competition in some economic sectors (for example, agriculture) and monopolistic competition in others sectors (for example, manufactured industry). These assumptions explain the existence of "backward and forward" linkages that create growth forces centripetal (having underlying monopolistic competition and increasing returns to scale) and centrifuges forces (because there are sectors in perfect competition with constant returns to scale). To verify these forces and linkages there will inevitably mobility of factors, including labor. Generally, the result of these links, forces and labor mobility is the formation of structures central-periphery, with benefits for the richest and prosperous.

Therefore, with this context, it appears that the current trend of various economic theories is to consider that the labor mobility accentuates regional disparities. Even writers in the line of neoclassical theory, as Barro and (6)Sala-i-Martin (1991), associated with endogenous growth theory, now admit that the mobility of labor reacts to processes of convergence and reduce regional disparities, but only if some conditions are met. That is, left to disappear the idea of absolute convergence for the same "steady state" of neoclassical influence, to a perspective of conditional convergence for differents "steady states".

## 2. THEORETICAL MODELS

We consider here, the models related to the migratory balance of (7)Salvatore (1977), (8)Katseli et al. (1989) and (9)Soukiazis (1995) and the models of the New Economic Geography of (10)Epifani et al. (2005). The choice of these models has to do with the fact that seem to be more closely aligned with the objectives set for this work initially just in the abstract. That is, models Salvatore (1977), Katseli et al. (1989) and Soukiazis (1995) are models simpler and can identify the determinants of labor mobility and the Epifani et al. (2005) is a more complete model that allows us to analyze the dynamics associated with the spatial evolution with implications for labor migration and unemployment.

### 3. THE MODEL USED

The model estimated in this study is what is presented below in Box 1. Are represented in the model presented below in Box 1 some new factors, mentioned in the economic theory, such as the effects of congestion, through the availability of housing.

**Box 1**: Balances migration as a function of economic factors and basic equipment (amenities)

$$(SM/PA)_t = c_0 + c_1(r_I - r_E)_t + c_2(D_I - D_E)_t + c_3(A_I)_t + c_4(s_I - s_E)_t + c_5(f_I - f_E)_t \quad (1)$$

SM/PA = net migration from one country or region with the outside, as a percentage of total active population of the country or region;
$r_I$-$r_E$ = difference between the growth rates of real output, with $r_I$ to be the annual growth rate of real output of the originating country or region and $r_E$ being the average growth rates of real GDP in all countries or regions destination;
$D_I$-$D_E$ = difference between the internal unemployment rate and the external average;
$A_I$ = number of employees in agriculture of the country or region of origin;
$s_I$-$s_E$ = difference between the internal growth rate of wage and external average;
$f_I$-$f_E$ = difference between the internal growth rate of housing and external average.

**Box 2**: An alternative model of net migration with spatial effects

$$(SM/PA)_t = c_0 + \rho(W(SM/PA)) + c_1(r_I - r_E)_t + c_2(D_I - D_E)_t + c_3(A_I)_t + c_4(f_I - f_E)_t + \varepsilon \quad (1)$$

W = matrix of distances;
$\rho$ = autocorrelation coefficient (the component "spatial lag");
$\varepsilon$ = error term (the component "spatial error", and $\varepsilon = \lambda W\varepsilon + \xi$).
The other variables and coefficients have the same meaning as that before.

In the estimates with spatial effects there are some spatial econometric techniques that are commonly used. In particular, the Moran's I statistic that is used to identify the existence of local and global spatial autocorrelation, the strategies of specification classical in six steps of (11)Florax et al. (2003) and LM tests to identify which form is most appropriate to the model specification, in other words, the component "spatial lag" (where the dependent variable is spatially lagged through the matrix W), or the component with the "spatial error "(where is the error term is spatially lagged).

### 4. THE DATA

The statistical information collected in the statistics of the INE (2006) and is relative to the variables of the models presented in Box 1 and 2, for the NUTS II (1996-2002) and NUTS III (1991 and 2001). There was a concern of do not join the data from 2003 and 2004 with the others from the previous years, because there have been changes since 2003 in the distribution of the various NUTS III NUTs II.
Then it will proceed to the analysis of the data, first at the level of NUTS II and later at the level of NUTS III.

### 5. EMPIRICAL EVIDENCES

Then we present empirical evidence for the different NUTS II, from 1996 to 2002, and for the NUTS III in 1991 and 2001. The estimation methods used are the fixed effects and random effects, with panel data, for the estimates made at the level of NUTS II, and OLS and maximum likelihood, with "cross section" data, in the estimates made at NUTS III. The consideration of the OLS and maximum likelihood estimates in the "cross section" for the NUTS III, has to do with the usual procedures in the estimations with spatial effects. That is, first estimates the OLS model to identify the existence or not of spatial effects and subsequently, in the case of identifying spatial effects, the model is estimated with the method of maximum likelihood.

### 51. EMPIRICAL EVIDENCES ON THE LEVEL OF NUTS II

Analyzing the results presented below in Table 1 for the estimation of equation (1) Box 1, we verify which the estimation method which we must to take in count is that of random effects, given the value of the Hausman test (no significant statistics). On the other hand, only the coefficients associated with the relative growth rates of real output, unemployment rates and the relative share of agricultural employment are that have statistical significance. The first coefficient referred has positive effect (only significant for 10%) and the last two negative effects (as it was expected, given the theory). It should be noted, however, that the coefficient associated with the share of employment has the highest marginal effect (-1.913).

For these reasons, we conclude that the regional mobility of labor in mainland Portugal is positively affected by growth rates of real output, in other words, greater is the difference between the rate of growth of real output of a region and the average growth rates of other regions most is the migration of workers into the region. On the other hand, it appears that mobility is negatively related to unemployment rates and the relative share of agricultural employment. That is, higher the unemployment rate of a region and greater the weight of the agricultural sector, lower is the labor migration to this region.

The growth rates for wages and growth rates on the housing stock does not have statistical significance and because this they have no influence on national labor mobility. What is not a surprising, given the Portuguese regional context.

**Table 1**: Results of panel estimations, with the equation of net migration for the NUTS II in the period 1996-2002

$$(SM/PA)_t = c_0 + c_1(r_I - r_E)_t + c_2(D_I - D_E)_t + c_3(A_I)_t + c_4(s_I - s_E)_t + c_5(f_I - f_E)_t$$

|      | $c_0$ | $c_1$ | $c_2$ | $c_3$ | $c_4$ | $c_5$ | G.L. | $R^2$ | SEE | T.H. |
|------|-------|-------|-------|-------|-------|-------|------|-------|------|------|
| LSDV | (#) | 0.235 (1.062) | -0.008** (-1.890) | -0.746* (-2.228) | -0.027 (-0.086) | 0.150 (0.618) | 20 | 0.693 | 0.013 | 6.157 (0.188) |
| GLS  | 0.148* (2.627) | 0.310** (1.802) | -0.020* (-3.234) | -1.913* (-3.153) | -0.078 (-0.333) | 0.247 (1.395) | 18 | 0.708 | 0.013 | |

**Note: LSDV, method of estimation with fixed effects; GLS estimation method with random effects; * Coefficient statistically significant at 5%; ** Coefficient statistically significant at 10%; GL, Degrees of freedom; SEE, standard deviation estimation; TH, Hausman Test; (#), all "dummies" statistical significance and values are very close. Figures in brackets are the t-statistics.**

### 5.2. EMPIRICAL EVIDENCES ON THE LEVEL OF NUTS III

Table 2 shows the results of the estimations, with the OLS estimation method, of the equation of net migration (Box 2), at the level of NUTS III of Portugal, and for the years 1991 and 2001 (years that correspond to the Portuguese Census and are unique for demographic statistics with a finer spatial disaggregation). The equation was modified by removal of the variable on wages, since there are no data.

The estimation results confirm that there is no spatial autocorrelation, "spatial lag" or "spatial error" (since the LM tests have no statistical significance) for net migration/population active, and show that for the level of NUTS III and for years considered the evolution of net migration is explained solely by the availability of housing. The positive sign of the coefficient (as expected) means that higher the rate of growth in the number of houses in a region compared with the average of other regions, increased migration of labor to the region. The fact that there is no autocorrelation "spatial lag" or "spatial error" means that the migration balance or are not influenced by net migration or by other factors of the neighboring regions, respectively.

Interestingly, the relative growth of basic equipment ("amenities") has no statistical significance at the level of NUTS II and is the only variable being significant at the level of NUTS III, which may have to do with the level of geographical disaggregation. Probably because the mobility depends on a more aggregated level of professional opportunities, while the mobility to a lower level depends on the availability of housing.

**Table 2**: Results of OLS estimates with "cross section" data subject to spatial effects, with the equation of net migration for the NUTS III and in the years 1991 and 2001

$$(SM/PA)_t = c_0 + \rho(W(SM/PA)) + c_1(r_I - r_E)_t + c_2(D_I - D_E)_t + c_3(A_I)_t + c_4(f_I - f_E)_t + \varepsilon$$

|     | $c_0$ | $c_1$ | c2 | c3 | c4 | JB | BP | KB | M'I | $LM_l$ | $LMR_l$ | $LM_e$ | $LMR_e$ | $R^2$ | SEE |
|-----|-------|-------|-----|-----|-----|-----|-----|-----|------|--------|---------|--------|---------|-------|------|
| OLS | 0.003 (0.170) | 0.048 (1.448) | -0.011 (-0.040) | -0.169 (-0.295) | 0.155* (2.165) | 5.061 | 7.491 | 4.751 | 1.975 | 0.034 | 1.018 | 0.619 | 1.602 | 0.201 | 0.025 |

**Note: JB, Jarque-Bera test for normality; BP, Breusch-Pagan test for heteroscedasticity; KB, Koenker-Bassett test for heteroskedasticity; M'I, Moran's I; LML, LM test for the component "spatial lag"; LMRL, robust LM test for the component "spatial lag"; LME, LM test for the component "spatial error"; LMRE, robust LM test for the component "spatial error"; *, statistically significant to 5%; **, statistically significant at 10%; SEE, standard deviation of the estimation.**

### 6. CONCLUSIONS

After the analysis of migration in Portugal, through the alternative model developed by Soukiazis (1995) and modified by us with the introduction of congestion effects (many of the developments cited in the New Economic Geography), using as "proxy" the housing stock (following procedures of (12)Hanson (1998) and (13)Antolin et al. (1997)), it is concluded that regions with higher unemployment rates and higher employment in agriculture are those that attract less people.

On the other hand, at the level of NUTS III is the housing stock (number of houses) which affects the mobility of populations. It was concluded, yet, that although there is spatial autocorrelation in terms of overall net migration is not enough to explain their evolution between the different NUTS III.

It is noted also that in the period 1996 to 2002, the Algarve was the region with higher percentages for net migration. Different trend showed the Alentejo region which has even negative migration balance at the beginning of the period, which is understandable, since it is the region with the highest rates of unemployment and highest percentage of employment in agriculture.

Based on Census 1991 and 2001, at the NUTS III level, it appears that the Alentejo Interior lose population and the coast and Algarve wins. Something similar we can see for the difference between the internal growth rates and external average, for the product, and the number of houses. Evolutionary trends almost inverse follow the average unemployment rates and average farm employment. There are however some cases it is worth noting, particularly the Grande Lisboa which usually in the variables analyzed does not follow the trend of the other regions of the coast, thus showing some signs of congestion.